\pdfoutput=1

\documentclass[11pt]{article}

\usepackage[final]{acl}

\usepackage{times}
\usepackage{latexsym}

\usepackage{subcaption}
\usepackage{amsmath, amsfonts,amssymb}
\usepackage{booktabs}
\usepackage{tabularx}
\usepackage{multirow}
\usepackage{MnSymbol}
\usepackage{array}
\usepackage{cleveref}
\usepackage{verbatim}
\crefformat{section}{\S#2#1#3}
\usepackage{diagbox}

\usepackage[linesnumbered,ruled]{algorithm2e}
\usepackage{makecell}
\usepackage{kotex}
\usepackage[T1]{fontenc}
\usepackage[utf8]{inputenc}
\usepackage{microtype}
\usepackage{inconsolata}
\usepackage{graphicx}
\newcommand{\len}[1]{\left|#1\right|}
\usepackage{ifthen}
\newcommand{\tr}[1]{{\color{red}#1}}
\newcommand{\cw}[1]{{\color{blue}#1}}
\usepackage{float}
\usepackage{tikz}
\newcommand{\blacksolid}{\raisebox{2pt}{\tikz{\draw[-,black,solid,line width = 1.0pt](0,0) -- (5.5mm,0);}}}
\newcommand{\blackdash}{\raisebox{2pt}{\tikz{\draw[-,black,dashed,line width = 1.0pt](0,0) -- (5.5mm,0);}}}

\title{Q-FAKER: Query-free Hard Black-box Attack via Controlled Generation }

\usepackage[symbol]{footmisc}

\author{CheolWon Na, YunSeok Choi\textsuperscript{\dag}, Jee-Hyong Lee\thanks{Co-corresponding authors.}\\
  College of Computing and Informatics\\
  Sungkyunkwan University\\
  \texttt{\{ncw0034, ys.choi, john\}@skku.edu}}

\makeatletter
\renewcommand{\@fnsymbol}[1]{\dag} 
\makeatother
\begin{document}
\maketitle
\begin{abstract}
Many adversarial attack approaches are proposed to verify the vulnerability of language models.
However, they require numerous queries and the information on the target model. Even black-box attack methods also require the target model's output information. 
They are not applicable in real-world scenarios, as in hard black-box settings where the target model is closed and inaccessible. 
Even the recently proposed hard black-box attacks still require many queries and demand extremely high costs for training adversarial generators.
To address these challenges, we propose \textbf{Q-faker} (\textit{\textbf{Q}uery-\textbf{f}ree H\textbf{a}rd Black-box Attac\textbf{ker}}), a novel and efficient method that generates adversarial examples without accessing the target model.
To avoid accessing the target model, we use a surrogate model instead. 
The surrogate model generates adversarial sentences for a target-agnostic attack. 
During this process, we leverage controlled generation techniques.
We evaluate our proposed method on eight datasets.
Experimental results demonstrate our method's effectiveness including high transferability and the high quality of the generated adversarial examples, and prove its practical in hard black-box settings.

\end{abstract}

\section{Introduction}
\label{section:introduction}

Language models have become crucial to various real-world applications
\citep{huang2024comprehensive, wang2024large}.
Despite their remarkable performance, these models are vulnerable to adversarial examples, such as word substitutions~\citep{papernot2016transferability, madry2018towards, choi2022tabs, nakamura-etal-2023-logicattack, huang2023robustness, burger-etal-2023-explanations}. 
Attackers can easily evade 
language models and achieve their malicious aims, such as spreading toxic content or rumors, by intentionally generating adversarial samples.
To address these issues, many studies have been conducted to analyze the adversarial vulnerability of language models.

Adversarial attack methods on language models can be categorized into white-box attacks, black-box attacks, and hard black-box attacks.
In white-box attacks, it is assumed that attackers have the internal structural information on the target model
\citep{guo-etal-2021-gradient, wang-etal-2022-semattack,liu-etal-2022-character, li2023white}. 
Black-box attacks do not require the internal information of the target model, but still require the output of the model, such as the logit scores or the predicted labels
\citep{deepwordbug, pwws, textfooler, sememepso, bert-attack, yu-etal-2022-texthacker, na-etal-2023-dip}. 
Hard black-box attacks do not require any information on the target model, such as its structure and outputs \citep{ATGSL, ct-gat}. It is assumed that
attackers do not have any internal information of the model, and lack access to the logit values or the predicted results of the target model.


\begin{table*}[t]
\centering
\small
\begin{tabular}{c|cc|cc}
\toprule

\multirow{3}{*}{\bf Attack Method} 
\rule{0pt}{2.5ex}
& \multicolumn{2}{c|}{\textbf{Requirements} ($\star$)} 
& \multicolumn{2}{c}{\textbf{Preparation Costs}} \\

& \multirow{1}{*}{\textit{Output}}
& \multirow{1}{*}{\textit{Predicted}}
& \multirow{1}{*}{\textit{Trainable}}
& \multirow{1}{*}{\textit{Extra}}  \\

& \multirow{1}{*}{\textit{Probability} }
& \multirow{1}{*}{\textit{Labels} }
& \multirow{1}{*}{\textit{Parameters}}
& \multirow{1}{*}{\textit{Train Dataset}} \\

\midrule
DeepwordBug (\citeauthor{deepwordbug}) & $\checkmark$ & $\checkmark$ & $-$ & $-$  \\
PWWS (\citeauthor{pwws}) & $\checkmark$ & $\checkmark$ &  $-$ & $-$  \\
TextFooler (\citeauthor{textfooler}) & $\checkmark$ &  $\checkmark$ & $-$ & $-$  \\
SememePSO (\citeauthor{sememepso}) & $\checkmark$ & $\checkmark$ & $-$ & $-$   \\
BERT-ATTACK (\citeauthor{bert-attack}) & $\checkmark$ &  $\checkmark$ & $-$ & $-$    \\
CT-GAT (\citeauthor{ct-gat})  & $-$ & $-$ & 406M (Enc.-Dec.) & Adversarial Datasets ($1.5M$)   \\
Q-faker (ours)  & $-$ & $-$ & 1K (single-layer) & Target Task Datasets ($0.2M$) \\

\bottomrule
\end{tabular}
\caption{Requirements for black-box attack methods. The star ($\star$) indicates features that necessitate access to the target model. Existing black-box methods require output information obtained from the target model, which limits their applicability in more restrictive black-box settings such as real-world scenarios.
}
\label{table:requirements}
\end{table*} 

In real-world applications, hard black-box attacks are necessary because
white-box and black-box attacks are impractical.
In most cases, neural network models operate within a system. The output of the system may be visible, but the model's own output is hardly accessible externally. 
Another requirement for real-world scenarios, is to minimize queries.
If attack methods try a large number of queries, they could easily be detected as suspicious actions. 
Since the target model is often unknown and resources are limited, hard black-box attacks must be model-agnostic and cost-efficient to be feasible in real-world applications.

Recently, hard black-box attack methods have been proposed by \citet{ATGSL} and \citet{ct-gat}. 
However, these methods present several limitations in real-world scenarios.
Their training relies on costly large-scale adversarial sentences obtained from various target models using various attack methods. Thus, their performance heavily depends on attack algorithms and target models which were used to obtain the datasets.
Despite using extensive datasets, they still require a large number of queries to achieve success, making them inefficient and impractical in real-world applications.
To address the challenges in real-world scenarios, we propose a novel hard-black box attack method, \textbf{Q-faker} (\textit{\textbf{Q}uery-\textbf{f}ree H\textbf{a}rd Black-box Attac\textbf{ker}}), that is query-free, cost efficient in training, and capable of target-agnostic attacks.
In the hard black-box setting, we have no access to the target model. 
To address this, we use a surrogate model instead of the target model to generate adversarial sentences for target-agnostic attacks.
For low-resource training, we build the surrogate model using a pretrained generative language model, which can approximate the capabilities of the target model.
We modify the model by adding a single-layer (classification head) consisting of 1K trainable parameters. 
For generating adversarial examples with zero queries, we utilize gradients from the surrogate model, and leverage controlled generation techniques~\citep{pplm}.

Table \ref{table:requirements} shows our contribution compared to existing methods. 
Compared to existing black-box attack methods, such as DeepwordBUG \citep{deepwordbug}, PWWS \citep{pwws}, TextFooler \citep{textfooler}, SememPSO \citep{sememepso}, and BERT-Attack \citep{bert-attack}), our proposed method does not require any access to the target model. 
  Additionally, our method incurs significantly lower training costs compared to CT-GAT \citep{ct-gat}, a hard black-box attack.

To verify the performance, efficiency and target-agnostic capability, we conduct various experiments on four real-world tasks, including disinformation, spam, toxic, and misinformation \citep{advbench}, 
across various target models, such as, BERT, XLNet, RoBERTa, DeBERTa, DistilBERT, and ALBERT.
Experimental results demonstrate the superiority of Q-faker not only over hard black-box baselines but also over black-box baselines which can access more information than Q-faker. 
We also confirm the high transferability of Q-faker through the experiments. Q-faker achieves higher and more consistent attack success rates across target models compared to the baselines.
The various experimental results are reported in \cref{section:experiment} and
\cref{section:analysis} including  overall performance, ablation study, transferability, quality evaluation of generated examples, and detector evasion performance.


\begin{figure*}[t]
\centering
    {\includegraphics[width=\textwidth]{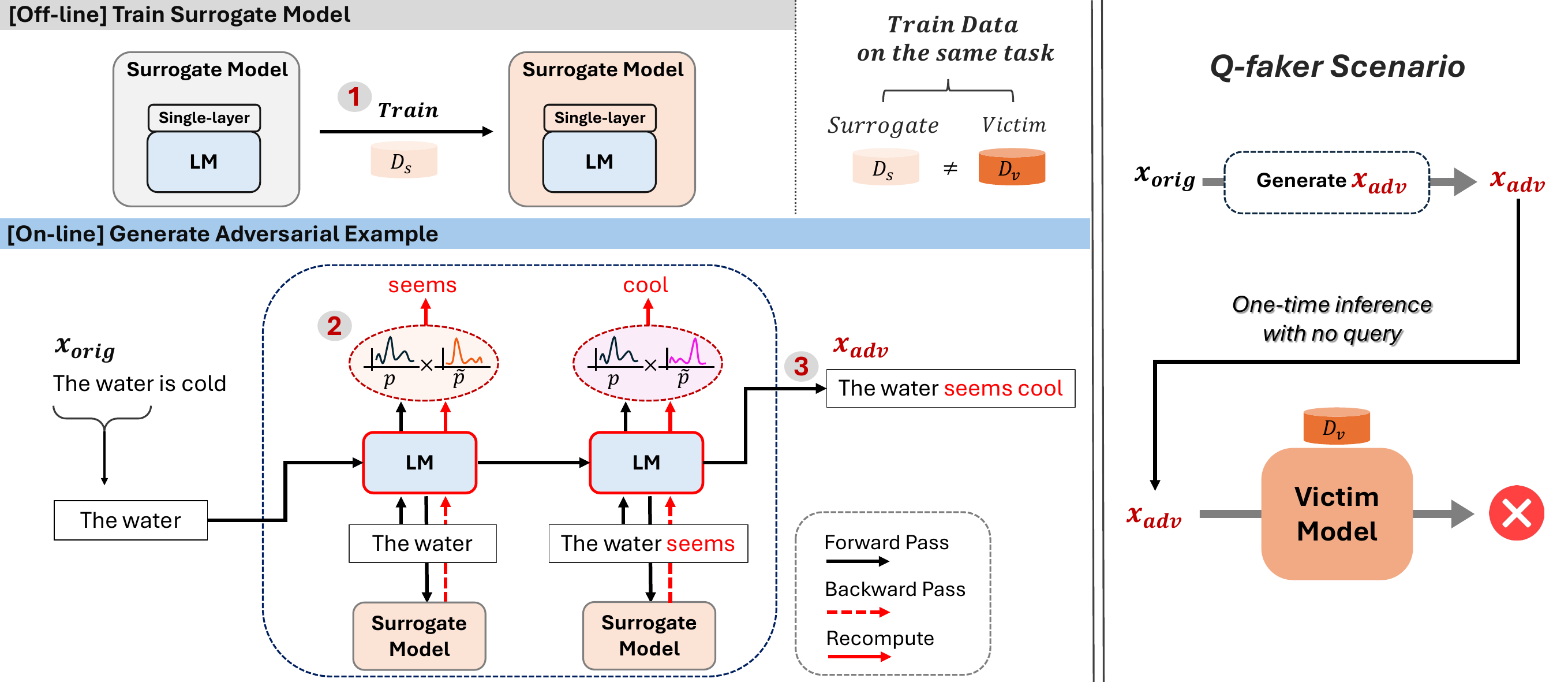}}\hfil
\caption{The process of Q-faker has three main steps: (1) Training the surrogate model using a different dataset for the same task as the target model; (2) Updating the language model using adversarial gradients from the surrogate model; and (3) Generating controlled adversarial examples from the updated language model. }
\label{figure:q-faker}
\end{figure*}


\section{Related Work}
\label{section:related_work}

The goal of adversarial attacks on language models differs whether the target model is NLU (Natural Language Understanding) or NLG (Natural Language Generation).
Adversarial attacks on NLU models, such as classification models, aim to mislead the output of model. 
Recent research about attacks on NLG models, such as generative LLMs (Large Language Models), have the objective of generating forbidden response or hallucinations like the jailbreaking \citep{chao2023jailbreaking, sun2024trustllm}.
The differences in attack methods are as follows: for NLU models, the approach involves manipulating the tokens of the input sentence. 
In generative LLMs, the approach includes adding additional prompts for various scenarios or tuning the parameters of soft prompts~\citep{perez2022red, zou2023universal, deng-etal-2023-attack, li-etal-2023-multi-step}.
This paper focuses on adversarial attacks on NLU models.
\paragraph{Black-box Attacks on NLU models}
The existing black-box attack methods are character or word manipulation approaches \citep{deepwordbug, pwws, textfooler, sememepso, bert-attack, yu-etal-2022-texthacker}. 
These methods rely on query-based algorithms and require a large number of queries to achieve a successful attack. 
They also require output information from the target model, which is not applicable in hard black-box setting as real-world scenarios. 
Recently, adversarial attack methods have been proposed that do not require any information about the target model \citep{ATGSL, ct-gat}. 
These methods train pre-trained language models on adversarial datasets that include adversarial examples generated by other attack algorithms. These approaches incur significant additional costs to obtain the adversarial datasets.

\section{Problem Statement} 
\label{section:problem_statement}

This study focuses on hard black-box attacks in real-world scenarios where queries and model's internal information are limited. 
Unlike our hard black-box setting, baselines are able to query the target model to obtain output information.

\paragraph{Hard black-box setting. } 
Since the our experimental setting is a hard black-box setting, we do not access to any information, including predicted labels and training dataset of target model. 
Therefore, the surrogate model's training dataset differs from the dataset used to fine-tune the target model.
Given the assumption that no output information from the target model is used, we generate only one adversarial sentence and conduct the test with a one attempt in the attack process.

\paragraph{Goal} Given an input sample ($s_i$, $y$), our goal is to find $s_{i}^{adv}$ by adding generated perturbation to $s_{i}$, that misleads the victim model.
The adversarial text $s_{i}^{adv}$, which has successfully attacked, needs to satisfy the followings criteria:
\begin{equation}
T(s_{i}^{adv}) \neq {y}, 
\end{equation}
where $T(s_{i}) = {y}$,
$T$ is the target model, and $y$ is the ground truth.

\section{Methodology} 
\label{section:methodology}
\begin{algorithm}[t]
\SetKwInOut{Input}{Input}
\SetKwInOut{Output}{Output}
  \Input{Original Example $s$ = [$x_{0},\dots,x_{n}$], Target Model $T$}
    \Output{Adversarial Example $s^{adv}$}
   \textbf{Initialization: }Begin with a pre-trained language model \text{LM} (in this paper GPT-2) and surrogate model  \text{SurModel} \\
   Train the SurModel on target task dataset \\
    $s$ $\leftarrow$ [$x_{0},\dots,x_{n}$] \\
    $s^{adv}$ $\leftarrow$ [$x_{0},\dots,x_{t}$], $t < n$  \\
    $maxlen$ $\leftarrow$ $\len{s}$ \\

    \While{$t < maxlen$}
    {
    $p(x_{t})$, $h_t$ $\leftarrow$ \text{LM}($x_{\leq t-1}$, $h_{t-1}$) \\ 
   $\mathcal{L}_{\text{sur}} \leftarrow \text{SurModel}({p}(x_t), \tilde{y})$ \\
   $\tilde{h_t} \leftarrow h_t + \eta \nabla_{h_t} \mathcal{L}_{\text{sur}}$ \\
   $\tilde{p}(x_{t+1})$, $h_{t+1}$ $\leftarrow$ \text{LM}($x_{\leq t}$, $\tilde{h_t}$) \\  
   ${\tilde{x}_{t+1}} \sim \tilde{p}_{fusion}(x_{t+1})$ \textcolor{teal}{\it \# See Section 4.3}\\
   $s^{adv}$ $\leftarrow$ [$x_{0},\dots,{\tilde{x}_{t+1}}$] \\
    }
    \If{$T(s) \leftarrow y$ and $T(s^{adv}) \neq y$} 
    {
        \textcolor{teal}{\it \# Attack Success} \\
        Return: $s^{adv}$ 
    }
    \caption{Q-faker Pseudo-code}
\label{algorithm:Q-faker}
\end{algorithm}

\begin{table*}[t]
\centering
\small
\begin{tabular}{c|cccc|cccc}
\toprule

\multirow{1}{*}{\bf Task} 
& \multicolumn{4}{c|}{\bf Spam}
& \multicolumn{4}{c}{\bf Sensitive Information} \\
\cline{1-9}
\rule{0pt}{2.5ex}
\multirow{2}{*}{\diagbox{Method}{Dataset}}
& \multicolumn{2}{c}{Assassin} 
& \multicolumn{2}{c|}{Enron} 
& \multicolumn{2}{c}{EDENCE} 
& \multicolumn{2}{c}{FAS} \\

& \multicolumn{1}{c}{\bf ASR (\textuparrow)} 
& \multicolumn{1}{c}{\bf Query (\textdownarrow)} 
& \multicolumn{1}{c}{\bf ASR (\textuparrow)} 
& \multicolumn{1}{c|}{\bf Query (\textdownarrow)} 
& \multicolumn{1}{c}{\bf ASR (\textuparrow)} 
& \multicolumn{1}{c}{\bf Query (\textdownarrow)} 
& \multicolumn{1}{c}{\bf ASR (\textuparrow)} 
& \multicolumn{1}{c}{\bf Query (\textdownarrow)}  \\

\midrule
DeepwordBug  & 0.0 & $\infty$ & 0.0 & $\infty$ & 2.5 & 9.00 & 3.4 & 8.82  \\
PWWS  & 0.0 & $\infty$ & 0.0 & $\infty$ & 0.0 & $\infty$ & 0.0 & $\infty$  \\ 
TextFooler  & 0.0 & $\infty$ & 0.0 & $\infty$ & 0.0 & $\infty$ & 0.0 & $\infty$  \\
SememePSO  & 0.1 & \underline{5.00} & 0.0 & $\infty$ & 0.6 & \underline{7.00} & 1.4 & \underline{8.29} \\
BERT-Attack  & 0.0 & $\infty$ & 0.1 & \underline{9.00} & 0.9 & 7.10 & 2.5 & 8.50 \\ 
CT-GAT & \underline{1.1} & 9.94 & \underline{0.2} & 9.99 & \underline{4.1} & 9.76 & \underline{10.0} & 9.32 \\
Q-faker (ours) & \textbf{5.1} & \textbf{0.00} & \textbf{0.7} & \textbf{0.00} & \textbf{54.6} & \textbf{0.00} & \textbf{43.3} & \textbf{0.00}  \\

\midrule
Clean Acc.  
& \multicolumn{2}{c}{98.4\%} 
& \multicolumn{2}{c}{99.6\%}
& \multicolumn{2}{c}{95.9\%}
& \multicolumn{2}{c}{97.4\%}\\
\bottomrule
\end{tabular}

\begin{tabular}{c|cccc|cccc}
\toprule

\multirow{1}{*}{\bf Task} 
& \multicolumn{4}{c|}{\bf Disinformation}
& \multicolumn{4}{c}{\bf Toxicity} \\
\cline{1-9}
\rule{0pt}{2.5ex}
\multirow{2}{*}{\diagbox{Method}{Dataset}}
& \multicolumn{2}{c}{CGFake} 
& \multicolumn{2}{c|}{Amazon-LB} 
& \multicolumn{2}{c}{Jigsaw} 
& \multicolumn{2}{c}{HSOL} \\

& \multicolumn{1}{c}{\bf ASR (\textuparrow)} 
& \multicolumn{1}{c}{\bf Query (\textdownarrow)} 
& \multicolumn{1}{c}{\bf ASR (\textuparrow)} 
& \multicolumn{1}{c|}{\bf Query (\textdownarrow)} 
& \multicolumn{1}{c}{\bf ASR (\textuparrow)} 
& \multicolumn{1}{c}{\bf Query (\textdownarrow)} 
& \multicolumn{1}{c}{\bf ASR (\textuparrow)} 
& \multicolumn{1}{c}{\bf Query (\textdownarrow)}  \\

\midrule
DeepwordBug & 0.0 & $\infty$ & 0.6 & \underline{6.67} & 8.1 & 8.05 & 7.1 & 7.63 \\
PWWS & 0.0 & $\infty$ & 0.0 & $\infty$ & 0.0 & $\infty$ & 0.2 & 9.00 \\ 
TextFooler & 0.0 & $\infty$ & 0.0 & $\infty$ & 0.0 & $\infty$ & 0.2 & 8.00 \\
SememePSO & 0.0 & $\infty$ & 0.1 & 9.00 & 1.0 & \underline{7.10} & 4.6 & 7.09 \\
BERT-Attack & 0.0 & $\infty$ & 0.5 & 7.00 & 3.3 & 7.88 & 5.8 & 8.12  \\ 
CT-GAT & \underline{12.4} & \underline{8.38} & \underline{8.1} & 9.47 & \underline{32.4} & 7.76 & \textbf{55.8} & \underline{5.73} \\
Q-faker (ours) & \textbf{13.4} & \textbf{0.00} & \textbf{8.6} & \textbf{0.00} & \textbf{38.2} & \textbf{0.00} & \underline{53.1} & \textbf{0.00} \\

\midrule
Clean Acc.  
& \multicolumn{2}{c}{97.8\%} 
& \multicolumn{2}{c}{91.6\%}
& \multicolumn{2}{c}{92.5\%}
& \multicolumn{2}{c}{95.5\%}\\
\bottomrule
\end{tabular}

\caption{Comparison of our proposed method with the baseline methods on eight victim models. 
The results are based on a real-world scenario with a maximum query limit set to 10.
The best performance is in \textbf{boldface}, and the second is \underline{underlined}.
}
\label{table:main_result}
\end{table*} 
In this section, we provide a detailed explanation of Q-faker.
As shown in Figure \ref{figure:q-faker}, our proposed method consists of three main steps:
(1) Train the surrogate model to derive adversarial gradients for adversarially updating the output distribution of the generator. 
Since the surrogate model is trained on same target task dataset, we can generate examples with high transferability.
(2) Update the output distribution of the generator based on the adversarial gradients obtained from the surrogate model.
In this step, we utilize the generative pre-trained language model GPT-2 as the generator to generate adversarial examples that ensure fluency and grammatical correctness.
(3) Generate an adversarial example from the updated output distribution of the generator. 
During the attack process, we use only the single adversarial example generated by Q-faker. 
Thus, we achieve a query-free and applicable approach in hard black-box settings.
The proposed method is summarized in Algorithm \ref{algorithm:Q-faker}.

\paragraph{Controlled Generation} 
Controlled generation is a method, generating sentences with a specific objective. These methods are commonly used in various tasks such as intent-based text generation, text style transfer, and etc.
To the best of our knowledge, we are the first to adopt controlled generation for generating adversarial examples. We utilize a simple and efficient controlled generation method proposed by \citet{pplm}.

\subsection{Surrogate Model}
\label{subsection:surrogate_model}
In order to adversarially update the output of the generator to the target model, we use a surrogate model that has been trained on the same target task but with a different dataset from the one used to train the target model.
We utilize the GPT-2 as a LM of the surrogate model for all experiments.
The architecture of surrogate model consists of generator with a single-layer head that has 1k parameters. 
When training the surrogate model, only the single layer is trained while the generator is frozen.
We train the surrogate model to minimize the loss $\mathcal{L}_{sur}$ as follows: 
\begin{equation}
\mathcal{L}_{sur} = - \sum_{i=1}^{N} y_i \log(\hat{y}_i) 
\end{equation}

\subsection{Adversarial Output Distribution}
\label{subsection:update_dist}
In order to adversarially update the output distribution of the Language Model (LM), we adopt a controlled generation approach \citep{pplm}. We compute adversarial gradients from a surrogate model in \cref{subsection:surrogate_model}. 
The adversarial gradient is calculated to maximize the loss function, thereby guiding the model towards incorrect predictions.
Based on the obtained adversarial gradients, we adversarially update the output distribution of LM.
Let $h$ be the key-value pair of LM. We then update $h$ to shift the output distribution for adversarial generation. 
To obtain the updated $\tilde{h}$, we compute the adversarial gradient of the surrogate loss as follows:
\begin{equation}
    p(x_t), h_t = \text{LM}.\text{forward}(x_{<t}, h_{t-1})
\end{equation}
\begin{equation}
   \nabla_{h_t} \mathcal{L}_{sur} = \frac{\partial \mathcal{L}_{sur}}{\partial p(x_t)} \cdot \frac{\partial p(x_t)}{\partial h_t}
\end{equation}
\begin{equation}
   \tilde{h}_t \leftarrow h_t + \alpha \frac{\nabla_{h_t} \mathcal{L}_{sur}}{\| \nabla_{h_t} \mathcal{L}_{sur} \|^{\gamma}} 
\end{equation}
where $\alpha$ is the step size, and $\gamma$ is the scaling coefficient for the normalization term. we use 0.06 and 1.0, respectively. This update step is repeated 10 times. Then, we adversarially update the output distribution of LM with updated $\tilde{h}$ as follows:
\begin{equation}
    \tilde{p}(x_{t+1}), h_{t+1} = \text{LM}.\text{forward}(x_{\leq t}, \tilde{h}_t)
\end{equation}
This step is described in Algorithm \ref{algorithm:Q-faker} (line 7-10).
\subsection{Controlled Adversarial Example Generation}
\label{subsection:generation}
In order to preserve the meaning of the original sentence, half of it is used as given tokens. These tokens are then used as input to the LM.
For controlled generating of adversarial examples, we use the updated LM as described in \cref{subsection:update_dist}. 
To ensure the LM's fluency capability, we utilize the \textit{post-norm fusion} \citep{stahlberg-etal-2018-simple}. This process maintains the fluency of the unmodified language model (in this study, GPT-2). The next tokens are then sampled from the fusion distribution as follows:
\begin{equation}
\tilde{p}_{\text{fusion}}(x_{t+1}) = p_{\text{orig}}(x_{t+1})^{1-\lambda} \cdot \tilde{p}(x_{t+1})^{\lambda}
\end{equation}
As $\lambda$ approaches 1, this converges to the distribution from the updated LM, and as approaches 0, it converges to the unmodified LM's distribution. We set $\lambda$ to 0.97 in all experiments.

\section{Experiments}
\label{section:experiment}

In this section, we demonstrate the effectiveness of our method. We use the Advbench dataset, a security-oriented adversarial NLP benchmark that includes tasks related to real-world harmful content problems~\citep{rocket}.
More details of datasets and target models are provided in Appendix \ref{sec:appendix_a}.
\subsection{Tasks and Datasets}
We conduct experiments on four detection tasks from the Advbench benchmark: disinformation, toxicity, spam, and sensitive information. Each task consists of two datasets, resulting in a total of eight victim models trained on these datasets. We randomly selected 1,000 test samples from each dataset. For a fair comparison, we used the same random seed as in previous studies.
\subsection{Evaluation Metrics}
We use two metrics to evaluate the attack \textit{efficiency} of our method. To measure the \textit{quality} of the generated adversarial examples, we use also three metrics. 
The performance results reported in this paper represent the average of successful attack instances. 


\paragraph{Attack efficiency.}
We evaluate the efficiency of attack methods using the attack success rate and query time. (1) Attack Success Rate is the success ratio of attacks (\textbf{ASR}); the higher the ASR, the better the performance of an attack method. (2) The query time is defined as the number of queries required to succeed attacks (\textbf{Query}). 

\paragraph{Attack quality.}
To measure the quality of the generated adversarial examples, we use the following three metrics: 
(1) The cosine similarity of the Universal Sentence Encoder vector \citep{cer2018universal}, which measures the semantic similarity between the original sentence and adversarial sentence (\textbf{USE}); 
(2) Perplexity of language model, which assesses the fluency of the generated sentence (\textbf{PPL});
and (3) ({\bf $\Delta$I}) indicates the increase in grammatical errors.
\subsection{Implementation}
\paragraph{Victim model.} We use BERT, the most representative pre-trained language model in the NLU task. We fine-tune a separate victim model for all datasets, building a total of eight victim models. 
We use the same split of the training dataset as \citet{ct-gat} did.

\paragraph{Baselines.}
For a fair comparison, all baselines were re-implemented using the same fixed seed. 
We used the NLP attack package, OpenAttack \citep{openattack}, to implement some baselines.
We select high-representative attack methods at both the character-level and the word-level.
At the character-level, we select TextFooler \citep{textfooler}, PWWS \citep{pwws}, and DeepwordBug \citep{deepwordbug}, while at the word-level, we use BERT-attack \citep{bert-attack} and SememePSO \citep{sememepso}.
We also compare with a strong baseline, CT-GAT, which has high-transferability. 
To implement CT-GAT, we use the source provided by \citet{ct-gat} to reproduce the results. When we implement  CT-GAT, we train the pre-trained Encoder-Decoder model, BART. 

\paragraph{Hyperparameters.} 
Q-faker has two hyperparameters. 
$r$ is the ratio of given tokens to the length of the original sentence. 
It is a hyperparameter that determines how many of the original sentence is used as input to the generator for generating adversarial sentences. 
$\lambda$ is a hyperparameter used for post-norm fusion in \cref{subsection:generation}. The closer $\lambda$ is to 1, the output distribution of the generator converges to an adversarially updated distribution.
We set $r$ to 0.5 and $\lambda$ to 0.97 in all experiments.


\subsection{Experimental Setup}
%
Since the baselines require output information from the target models, we allow them to make iterative queries to the target models in a black-box setting, with 10 queries in Table \ref{table:main_result} and 20 queries in Figure \ref{figure:limited_query}, respectively.
On the other hand, we evaluate our method in hard black-box setting. 
We evaluate our method with only one inference, without any target model's information including train datasets.

\paragraph{Cross-dataset setting.} 
To adhere to the hard black-box setting, we conduct experiments using cross-dataset setting. 
We generate adversarial sentences using a dataset different from the target task dataset, because attackers cannot access the dataset for the target task.
For example, if the target model is trained with the Assassin dataset for spam detection, the surrogate model is trained on the Enron dataset.

We ensure that both models are independently trained with two distinct datasets, each independently collected from real-world sources.
Our method is validated in this setting on all the experiments in this paper.
%
This experimental design rigorously validates hard black-box setting. 

\begin{figure}[t]
\centering
	\begin{subfigure}[t]{0.495\textwidth}
		\includegraphics[width=\textwidth,height=0.14\textheight]{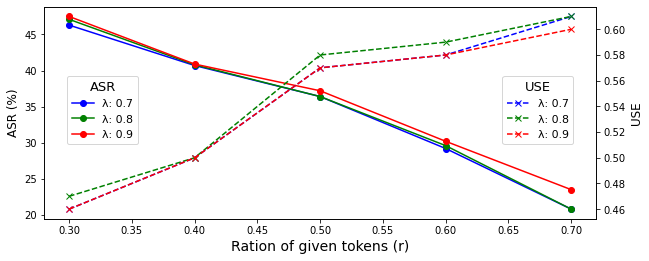}
	\end{subfigure}
\caption{Results of ablation study on Jigsaw. The solid lines (\protect\blacksolid) represent ASR (left y-axis), the dashed lines (\protect\blackdash) represent USE (right y-axis).}
\label{figure:ablation}
\end{figure}


\subsection{Experimental Results}
\label{subsection:experiment_results}
\paragraph{Main results.} 
Table \ref{table:main_result} shows the attack success rate and number of queries when the queries to the victim model are limited to 10. 
The results indicate that query-based methodologies mostly fail with the ASR close to zero.
As shown in Table \ref{table:main_result}, Q-faker outperforms the baselines in all metrics except ASR on HSOL. 
Since CT-GAT utilized the HSOL dataset during their training, it achieves an ASR approximately 2.7\%p higher than our method on the HSOL.
However, our proposed method, Q-faker, mostly shows the best on all datasets. 
Specifically, Q-faker impressively outperforms in the sensitive information task, with ASR differences ranging from at least 4 times to as much as 40 times higher compared to the baselines.

\paragraph{Ablation study.} 
We conduct an ablation study on the Jigsaw dataset to compare the effects of adversarial distribution ($\lambda$) and given tokens ($r$).
As shown in Figure \ref{figure:ablation}, increasing $\lambda$ slightly improves ASR. 
This indicates that adversarial distribution is helpful to attack, but not highly sensitive to $\lambda$.
When $r$ is reduced below 0.5, ASR increases while the USE drops significantly, resulting in substantial shift of the original sentence's meaning. 
This indicates a trade-off between ASR and USE based on $r$.
When $r$ is around 0.5, we effectively mitigate the trade-off between ASR and the preservation of the original meaning.

\section{Further Analysis}
\label{section:analysis}

\begin{figure*}[t]
\centering
	\begin{subfigure}[t]{0.24\textwidth}
        \caption*{\hspace{0.4cm} \small Enron}
		\includegraphics[width=\textwidth,height=0.16\textheight]{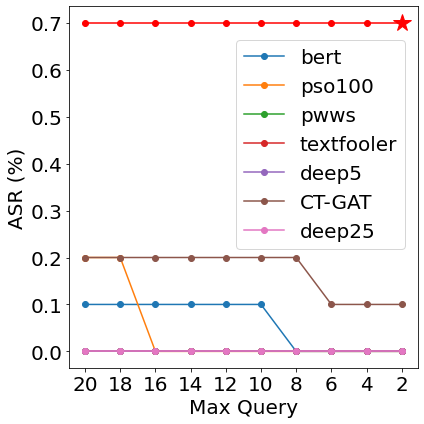}
	\end{subfigure}
    \begin{subfigure}[t]{0.24\textwidth}
        \caption*{\hspace{0.4cm} \small EDENCE}
		\includegraphics[width=\textwidth,height=0.16\textheight]{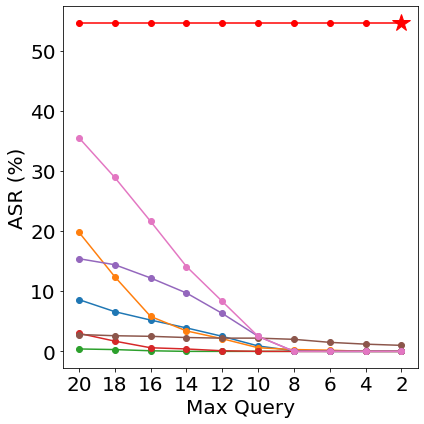}
	\end{subfigure}
    \begin{subfigure}[t]{0.24\textwidth}
        \caption*{\hspace{0.4cm} \small CGFake}
		\includegraphics[width=\textwidth,height=0.16\textheight]{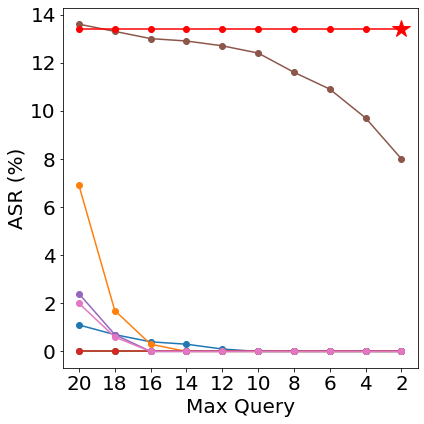}
	\end{subfigure}
    \begin{subfigure}[t]{0.24\textwidth}
        \caption*{\hspace{0.4cm} \small Jigsaw}
		\includegraphics[width=\textwidth,height=0.16\textheight]{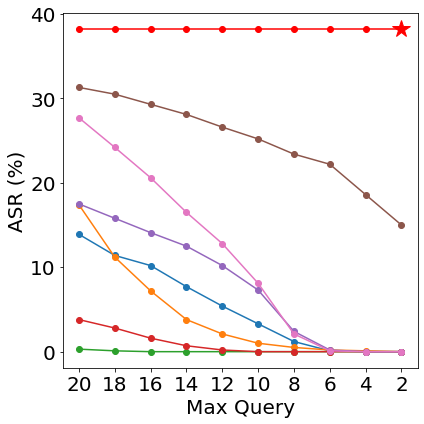}
	\end{subfigure}
\caption{
Comparison of ASR according to the number of queries (from 20 to 1). The red star(\textcolor{red}{ $\star$}) is our  method.
As the number of queries accessible to the target model becomes more restricted, the ASR of baseline methods drops to near zero. This demonstrates the superiority of our method in real-world scenarios with limited queries.
}
\label{figure:limited_query}
\end{figure*}

\begin{figure*}[t]
\centering
	\begin{subfigure}[t]{0.495\textwidth}
        \caption{Q-faker (ours)}
		\includegraphics[width=\textwidth,height=0.2\textheight]{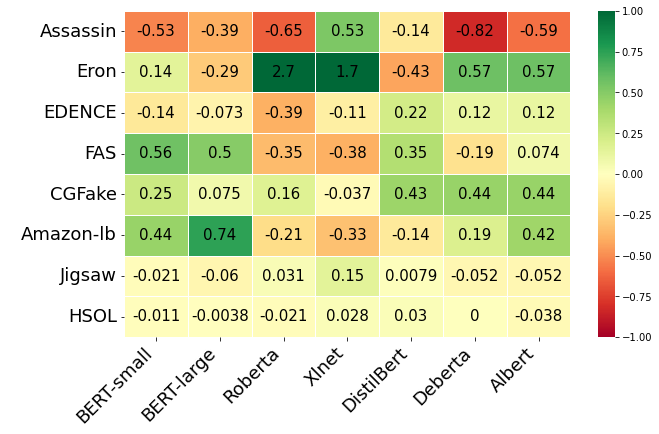}
	\end{subfigure}
    \begin{subfigure}[t]{0.495\textwidth}
        \caption{CT-GAT}
		\includegraphics[width=\textwidth,height=0.2\textheight]{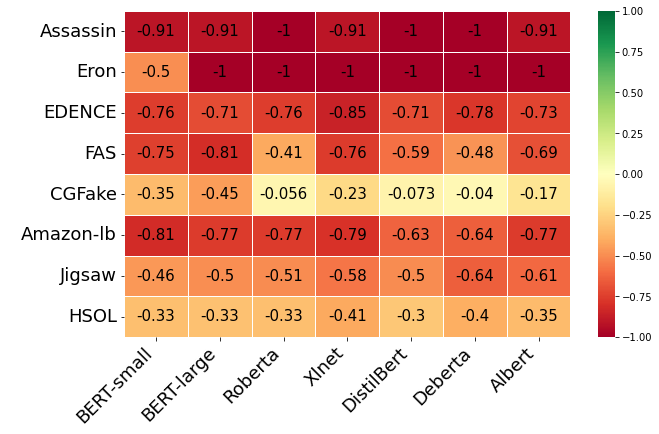}
	\end{subfigure}
\caption{Consistent attack capabilities on various target models. This heatmap illustrates the difference ratio in ASR between BERT-base and other models. We compare our method (left) with CT-GAT (right). Colors closer to green indicate higher ASR on other target models, highlighting the high transferability.}
\label{figure:transferability}
\end{figure*}

In this section, we further analyze the effectiveness of our proposed method through additional experiments. 
We demonstrate the superiority of Q-faker in various scenarios, including extremely limited queries, transferability, qualitative analysis and adversarial defenses. 
Additional results and more detailed, including time-complexity and feasibility of LLM attacks, are provided in Appendix \ref{sec:appendix_b}.

\subsection{Number of Queries} 
\label{subsection:limited_query}
We conduct further experiments under real-world scenarios where the number of queries is extremely limited. Figure \ref{figure:limited_query} shows the results of experiments conducted with different numbers of queries from 20 down to 1. 
When the number of queries is extremely limited to 1, our proposed method achieves significantly higher ASR compared to all other baseline methods across all datasets. 

\subsection{Transferability} 
\label{subsection:transferability}
To  validate the target-agnostic capability of our method, we conduct transferability experiments.
We calculate the ASR difference between the reference model and each of comparison models to assess how consistently our method performs across various models.
We choose the BERT-base as the reference model, and other pretrained language models as the comparison models, respectively.
Figure \ref{figure:transferability} is a heatmap showing the difference ratios in ASR between the reference model and the comparison models.
Cell values closer to zero signify more consistent performance across various models.
The positive cell value indicates that attack method has a higher ASR on the comparison model than the reference model, while the negative cell value indicates lower attack performance. 
For example, the cell value of (HSOL, BERT-small) is -0.011 which means that our method shows very similar performance  on both the comparison and the reference models.
In Figure \ref{figure:transferability}, our method (Q-faker) shows that most cells exhibit positive values or close to zero, whereas CT-GAT exhibits negative values in all cases.
The results show that our method generally consistent the target-agnostic attack capability across various target models, which indicates high transferability of our method.
\subsection{Generated Examples Quality Analysis}
We evaluate the quality of the adversarial examples generated by attack methods.
To validate a comprehensive quality analysis, we conduct additional the following experiments: automatic metrics, ChatGPT prompts evaluation, and human evaluation.
\paragraph{Automatic evaluation.} 
For a quantitative evaluation of quality, we conduct experiments using automated metrics as follows: \textit{USE}, \textit{PPL}, and $\Delta$\textit{I}.
Table \ref{table:result_quality} shows that our method show slightly lower \textit{USE} but better perform in \textit{PPL} and $\Delta$\textit{I}. 
The generator-based methods such as Q-faker and CT-GAT, have more differences in representation space compared to token manipulate-based methods. 
However, our proposed method shows notably lower perplexity and great grammatical correctness. 
These results highlight the high quality that appears natural to human judges.
\begin{table}[t]
\centering
\small
\begin{minipage}{\linewidth}
        \centering
        \begin{tabular}{l|ccc}
            \toprule
            Method & {\bf USE (\textuparrow)} & {\bf PPL (\textdownarrow)} & {\bf $\Delta$ I (\textdownarrow)} \\
            \midrule
            PWWS & \textbf{0.85} & 194.32 & 21.01 \\
            BERT-Attack & 0.78 & 1280.94 & 3.56 \\
            CT-GAT & 0.74 & 94.40 & 8.78 \\
            Q-faker (ours) & 0.75 & \textbf{25.95} & \textbf{-0.49} \\
            \bottomrule
        \end{tabular}
    \end{minipage}%
\caption{Comparison of adversarial examples generated by attack methods on Amazon-LB. Token manipulation approaches (PWWS and BERT-ATTACK)  have high \textit{USE} scores, whereas generation-based methods (CT-GAT and Q-faker) perform better in \textit{PPL} and $\Delta I$. 
}
\label{table:result_quality}
\end{table}

\begin{table}[t]
\centering
\small
\begin{minipage}{\linewidth}
        \centering
        \begin{tabular}{l|c|c}
            \toprule
            Metric & Q-faker (ours) & \makebox[1.5cm][c]{CT-GAT} \\
            \midrule
            Naturalness & \textbf{108} & 58 \\
            Meaning preservation & \textbf{95} & 71 \\
            Grammar & \textbf{112} & 54 \\
            \bottomrule
        \end{tabular}
    \end{minipage}%
\caption{ChatGPT prompt evaluation results. 
The results represent the total times the better generated sentence was chosen.
}
\label{table:chatgpt_eval}
\end{table}

\begin{table}[t]
\centering
\small
\begin{tabular}{cc|ccc}
\toprule
\multicolumn{2}{c}{\textbf{Dataset}} & \textbf{Accuracy} & \textbf{MP} & \textbf{Natural} \\
\midrule
\multirow{2}{*}{\textbf{FAS}} & Original & 0.84 & - & 3.93 \\
& Adversarial & 0.82 & 3.84 & 4.11 \\
\midrule
\multirow{2}{*}{\textbf{Assassin}} & Original & 0.97 & - & 3.64 \\
& Adversarial & 0.94 & 3.82 & 3.87 \\
\bottomrule
\end{tabular}
\caption{
Human evaluation results. Since the original datasets are online-data collected from the real-world, they mostly consist of informal, colloquial, and ungrammatical sentences. Therefore, adversarial sentences generated by our LM-based method have an advantage in naturalness.}
\label{table:human_eval}
\end{table}

\paragraph{ChatGPT prompt evaluation.} 
We conduct experiments using ChatGPT API to evaluate the quality of the generated sentences. 
Recent studies have raised fairness issues about human annotators and suggested that LLM-evaluators can be more reliable than human-evaluators \citep{chatgpthumanexperts}. 
In other lines of research, some studies address concerns about bias in LLM-evaluators \citep{bias-in-llms}.
To mitigate such biases, we utilize to the GPT-Rank template proposed by \citet{llm-blender}. 
For more details about prompt are described to the Appendix \ref{sec:appendix_c}. 
The experimental results show that our method outperforms on all metrics, as shown in Table \ref{table:chatgpt_eval}. 
Our method leverages a pre-trained language model on a large corpus, which is advantageous for naturalness (fluency) and grammatical correctness, and also better preserves the original semantic meaning. 

\paragraph{Human evaluation.} 
To further validate quality of generated sentences, we conduct a human evaluation to measure the semantic preservation and naturalness. 
We randomly select 50 pairs of original and adversarial texts from each of the FAS and HSOL datasets.
We ask four human evaluators the following three metrics: 
(1) \textbf{Accuracy}: the prediction of the label for the task (sensitive/toxic or not), (2) \textbf{MP} (Meaning Preservation): how well the adversarial sentence preserves the meaning of the original sentence, and (3) \textbf{Natural}: how natural the sentence appears, as if it were human-written without any manipulation. MP and Naturalness are scoring from 1-5 following \citep{bert-attack}.
Table \ref{table:human_eval} summarizes the results of the human evaluation, showing that our method effectively preserves the original meaning and increases naturalness.

\subsection{Adversarial Defense}
\label{subsection:defense}
Broadly, defense techniques s against adversarial attacks can be categorized into adversarial training~\citep{dong2021should, zhao2023generative} and detection~\citep{mosca-etal-2022-suspicious, zheng-etal-2023-detecting}.
In real-world scenarios, adversarial detection which preemptively blocks adversarial inputs, is more practical. To evaluate the detector evasion performance of the Q-faker, we conduct experiments to attack the detector on Amazon-LB datasets. 
We use the detector proposed by \citep{mosca-etal-2022-suspicious}.
Our proposed method demonstrates better performance compared to other approaches. 
The results are provided in Appendix \ref{appendix_subsection:defense}. 


\section{Conclusion}
\label{section:conclusion}
We proposed Q-faker, a novel, efficient, and query-free hard black-box attack method. 
Our method adopts controlled generation techniques to generate adversarial examples without any information from the target model. 
It demonstrates excellent performance without accessing target model information, and and has proven effective in real-world scenarios. 

\section*{Limitations}
We have some limitations about this work; (1) Our proposed method needs to know the specific task of the target model, (2) We do not consider scenarios where the query is infinite and access to the target model. However, our proposed method also provides room for an iterative attack to improve attack performance. 

\section*{Ethical Considerations}
We conduct experiments that are the security-oriented benchmark dataset, Advbench, which is open-source. We do not use any closed-source data, and our works ensure the ethical policy. 
However, the datasets used in this work contain potentially harmful content. We have chosen not to report directly on these harmful examples to consider ethical policy.
Our research focuses on adversarial attack methods that can be applied in real-world scenarios, this could potentially be misused by malicious users. They can spread rumors or spam emails. 
Therefore, the researcher in this related work must strictly adhere to ethical standards.


\section*{Acknowledgments}
This work was supported by Institute of Information \& communications Technology Planning \& Evaluation(IITP) grant funded by the Korea government(MSIT) (RS-2019-II190421, AI Graduate School Support Program(Sungkyunkwan University), 10\%), (IITP-2025-RS-2020-II201821, ICT Creative Consilience Program, 10\%), (No.RS-2021-II212068, Artificial Intelligence Innovation Hub, 10\%), (IITP-2025-RS-2024-00437633, ITRC(Information Technology Research Center), 70\%)

\bibliography{custom}

\clearpage
\appendix

\section{Experimental Setup}
\label{sec:appendix_a}
\subsection{Dataset} 
We utilize Advbench benchmark dataset. This dataset consists of data collected from real-world scenarios. The dataset statistics are presented in Table \ref{table:dataset_statistics}
\paragraph{Spam.} 
This task is to detect spam message including advertising, scams, phishing, and more.
This task is crucial for improving security and maintaining user trust.
\paragraph{Sensitive Information.} 
Detecting sensitive information in text is vital to prevent data leakage.
This task focuses on detecting sensitive information from companies, including intellectual property and product development updates and individuals information.
\paragraph{Disinformation.} 
The fake information is caused by subjectively facts.
This task is to identify deliberate fabrication of information as follows: (1) Artificial comments reversing the black and white; (2) Generated nonexistent information.
\paragraph{Toxic.} Malicious toxic texts are widespread in the web. The toxic texts detection is to identify for toxic contents including sexism, racism, cyber-bullying, and etc. 

\subsection{Surrogate and Target Models}
\paragraph{Surrogate Model.} In all experiments, we use GPT2-medium as the language model for generation, and we utilize a surrogate model by adding a classification head (a single layer) to GPT2 for task-specific training in order to obtain task-related gradients.

\paragraph{Target Models.} In this paper, BERT-base is used as the target model for the reported results. For Table \ref{figure:transferability}, we employ encoder-based models that demonstrate strong performance in classification tasks, including the BERT, Roberta, Xlnet, DistilBert, Deberta, and Albert. The performance of the fine-tuned target models for each task is reported in Table \ref{table:target_models}.
\begin{table}[t]
\centering
\small
\begin{tabular}{c|ccc}
\toprule

\multirow{1}{*}{Dataset}
& \multirow{1}{*}{$\#$ of Train}
& \multirow{1}{*}{$\#$ of Test}
& \multirow{1}{*}{Avg. Length} \\
\midrule
\multicolumn{4}{c}{Spam} \\ 
\midrule
Enron & 16159 & 7277 & 311.53  \\ 
Assassin & 2081 & 2066 & 308.50  \\ 

\midrule
\multicolumn{4}{c}{Sensitive Information } \\ 
\midrule
EDENCE & 51098 & 10328 & 21.79  \\ 
FAS & 33814 & 13294 & 29.27  \\ 

\midrule
\multicolumn{4}{c}{Disinformation} \\ 
\midrule
Amazon-LB & 17434 & 8522 & 100.13  \\ 
CGFake & 28290 & 12130 & 67.48  \\ 

\midrule
\multicolumn{4}{c}{Toxic} \\ 
\midrule
HOSL & 5832 & 2494 & 14.32  \\ 
Jigsaw & 30587 & 12180 & 58.42  \\ 

\bottomrule
\end{tabular}

\caption{Dataset statistics}
\label{table:dataset_statistics}
\end{table} 

\begin{itemize}
\item \textbf{BERT}~\cite{bert}, a one of the most widely used language models, known for its bidirectional context understanding.
\item \textbf{RoBERTa}~\cite{roberta}, refines BERT by removing the Next Sentence Prediction (NSP) objective to robustly optimization, and increasing the training duration, and utilizing a larger batch size and more data.
\item \textbf{XLNet}~\cite{xlnet}, a permutation language model without relying on masking, which improves its ability to model language relationships in more flexible and robust ways.
\item \textbf{DeBERTa}~\cite{deberta}, an advanced variant of BERT that improves upon BERT and RoBERTa by using a disentangled attention mechanism and enhanced decoding.
\item \textbf{DistilBERT}~\cite{distilbert}, a  faster and more efficient version of BERT, designed using knowledge distillation to retain.
\item \textbf{ALBERT}~\cite{albert}, a lightweight version of BERT designed to reduce the model size and training time by sharing parameters across layers and factorizing the embeddings.

\end{itemize}


\begin{table*}[t]
\centering
\small
\begin{tabular}{c|cc|cc|cc|cc}
\toprule

\multirow{2}{*}{\diagbox{Model}{Dataset}}
& \multicolumn{2}{c|}{\bf Spam}
& \multicolumn{2}{c|}{\bf Sensitive Information}
& \multicolumn{2}{c|}{\bf Disinformation}
& \multicolumn{2}{c}{\bf Toxic} \\
& \multicolumn{1}{c}{Assassin} 
& \multicolumn{1}{c|}{Enron} 
& \multicolumn{1}{c}{EDENCE} 
& \multicolumn{1}{c|}{FAS}
& \multicolumn{1}{c}{CGFake} 
& \multicolumn{1}{c|}{Amazon-LB} 
& \multicolumn{1}{c}{Jigsaw} 
& \multicolumn{1}{c}{HSOL} \\ 

\midrule
BERT & 98.4 \% & 99.6 \% & 95.9 \% & 97.4 \% & 97.8 \% & 91.6 \% & 92.5 \% & 95.5 \%  \\ 
RoBERTa  & 98.4 \% & 99.5 \% & 95.4 \% & 97.1 \% & 98.7 \% & 91.9 \% & 91.5 \% & 95.5 \%  \\
XLNet  & 98.8 \% & 99.6 \% & 95.4 \% & 96.5 \% & 97.9 \% & 91.7 \% & 91.5 \% & 95.6 \% \\ 
DeBERTa & 98.7 \% & 99.6 \% & 95.8 \% & 97.0 \% & 98.3 \% & 92.2 \% & 91.6 \% & 95.9 \% \\
DistilBERT & 98.3 \% & 99.3 \% & 95.6 \% & 97.5 \% & 98.1 \% & 91.2 \% & 90.8 \% & 95.9 \%  \\
ALBERT & 98.5 \% & 99.4 \% & 95.4 \% & 94.7 \% & 98.4 \% & 89.4 \% & 91.1 \% & 95.1 \%  \\
\bottomrule
\end{tabular}

\caption{Performance of fine-tuned target models.}
\label{table:target_models}
\end{table*} 

\section{Further Analysis}
\label{sec:appendix_b}


\subsection{Number of Queries}
\begin{figure}[t]
\centering
	\begin{subfigure}[t]{0.23\textwidth}
        \caption*{\hspace{0.4cm} \small Assassin}
		\includegraphics[width=\textwidth,height=0.16\textheight]{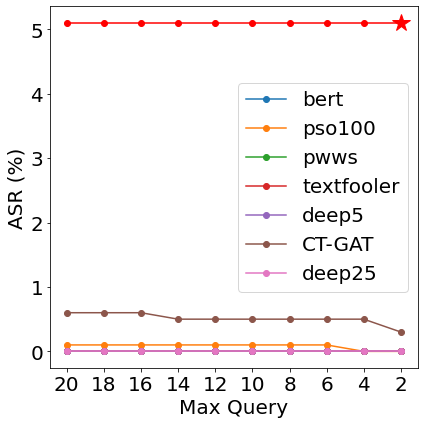}
	\end{subfigure}
    \begin{subfigure}[t]{0.23\textwidth}
        \caption*{\hspace{0.4cm} \small FAS}
		\includegraphics[width=\textwidth,height=0.16\textheight]{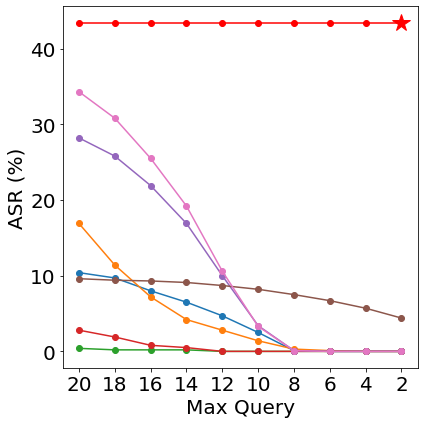}
	\end{subfigure}
    \begin{subfigure}[t]{0.23\textwidth}
        \caption*{\hspace{0.4cm} \small Amazon-LB}
		\includegraphics[width=\textwidth,height=0.16\textheight]{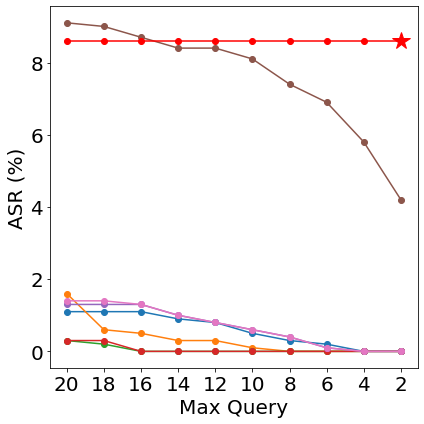}
	\end{subfigure}
    \begin{subfigure}[t]{0.23\textwidth}
        \caption*{\hspace{0.4cm} \small HSOL}
		\includegraphics[width=\textwidth,height=0.16\textheight]{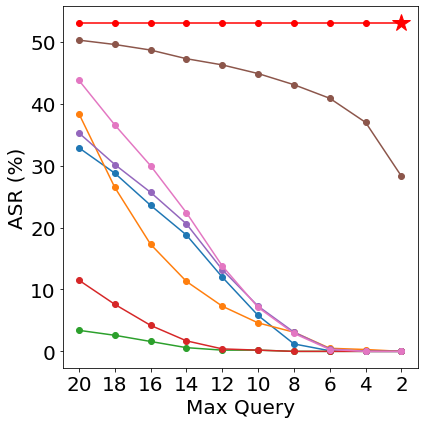}
	\end{subfigure}
\caption{
Comparison of ASR according to the number of queries (from 20 to 1). The red star(\textcolor{red}{ $\star$}) is our  method.
As the number of queries accessible to the target model becomes more restricted, the ASR of baseline methods drops to near zero. This demonstrates the superiority of our method in real-world scenarios with limited queries.
}
\label{appendix_figure:limited_query}
\end{figure}

We conduct same experiments in Section \ref{subsection:limited_query} on additional datasets. 
Figure \ref{appendix_figure:limited_query} shows the results of experiments conducted with different numbers of queries from 20 down to 1. 

\subsection{Transferability}
To additionally demonstrate the superiority of our method, we conduct experiments using target model trained from scratch, instead of pre-trained models.
In this experiment, we unlimit the number of queries for the baselines. We allow them to attack the source model until they succeed.
To conduct transferability experiments, we build a source model by training with different architectures, parameters, and datasets from those of the target model to be attacked.
Q-faker’s surrogate model is trained on the same data as the source model.
Table \ref{appendix_table:transferability} shows the ASR results when attacking target models using examples obtained from the source model. 
Since BERT-ATTACK utilizes BERT's MLM, they are advantageous when the target model belongs to the BERT family. Nevertheless, Q-faker mostly shows superior performance on all datasets.

\begin{table}[t]
\centering
\small
\begin{tabular}{l|cccc}
\toprule
\diagbox{Method}{Dataset} & Amzon. & HSOL & Assas. & FAS   \\
\midrule
\multicolumn{5}{c}{Source: BERT $\xrightarrow{}$ Target: CNN} \\ 
\midrule
TextFlooler & 4.43 & 37.00 & 4.89 & 6.80 \\
BERT-ATTACK & 3.96 & \textbf{49.40} & 17.72 & 33.80 \\
CT-GAT & 4.20 & 43.20 & 11.20 & 4.00 \\
\midrule
Q-faker (ours) & \textbf{6.39} & \textbf{49.40} & \textbf{23.50} & \textbf{64.60} \\
\bottomrule
\end{tabular}
\caption{Results of transferability in various cases with different types of source and target models. 
This setup involves different parameters, training data, and architectures. Since our method requires a generative surrogate model, we use GPT-2 as the surrogate model instead of the source model in all cases.
}
\label{appendix_table:transferability}
\end{table}

\begin{table}[t]
\centering
\small
\begin{minipage}{\linewidth}
        \centering
        \begin{tabular}{l|c|c}
            \toprule
            Metric & Q-faker (ours) & \makebox[1.5cm][c]{BERT-Att.} \\
            \midrule
            Naturalness & \textbf{102} & 64 \\
            Meaning Preservation & \textbf{84} & 82 \\
            Grammaticality & \textbf{92} & 74 \\
            \bottomrule
        \end{tabular}
    \end{minipage}%
\caption{Comparison with BERT-Attack using ChatGPT prompt. This experiment selects the better example generated by the two attack methods.
}
\label{appendix_table:chatgpt_eval}
\end{table}

\subsection{Automatic Evaluation}
We conduct experiments for evaluation quality on other datasets. 
Our generation-based method shows lower USE scores, which are representation-based, but it outperforms in more important quality metrics such as \textit{PPL} and \textit{$\Delta$ I}.
The results are reported in Table \ref{appendix_table:result_quality}.

\begin{table*}[t]
\centering
\small
\begin{tabular}{c|ccc|ccc|ccc}
\toprule

\multirow{2}{*}{\diagbox{Method}{Dataset}}
& \multicolumn{3}{c|}{CGFake} 
& \multicolumn{3}{c|}{HSOL} 
& \multicolumn{3}{c}{Jigsaw} \\

& \multicolumn{1}{c}{\bf USE (\textuparrow)} 
& \multicolumn{1}{c}{\bf PPL (\textdownarrow)} 
& \multicolumn{1}{c|}{\bf $\Delta$ I (\textdownarrow)} 
& \multicolumn{1}{c}{\bf USE (\textuparrow)} 
& \multicolumn{1}{c}{\bf PPL (\textdownarrow)}
& \multicolumn{1}{c|}{\bf $\Delta$ I (\textdownarrow)}  
& \multicolumn{1}{c}{\bf USE (\textuparrow)} 
& \multicolumn{1}{c}{\bf PPL (\textdownarrow)}
& \multicolumn{1}{c}{\bf $\Delta$ I (\textdownarrow)}  \\

\midrule

PWWS  
& 0.79 & 123.12 & 13.74
& 0.84 & 1433.60 & 2.83
& 0.82 & NaN & 4.87 \\ 

BERT-ATTACK 
& 0.79 & 291.22 & 13.28
& 0.76 & 393.87 & 3.43
& 0.83 & 1468.76 & 2.04 \\ 

CT-GAT
& 0.66 & 74.64 & 4.19
& 0.47 & 2.75 & 1.29
& 0.44 & 3.84 & 2.55  \\

Q-faker (ours)
& 0.55 & 45.33 & 0.64
& 0.53 & 204.31 & 0.2
& 0.57 & 138.86 & -1.69 \\

\end{tabular}

\begin{tabular}{c|ccc|ccc|ccc}
\toprule

\multirow{2}{*}{\diagbox{Method}{Dataset}}
& \multicolumn{3}{c|}{EDENCE} 
& \multicolumn{3}{c|}{Enron} 
& \multicolumn{3}{c}{Assassin} \\

& \multicolumn{1}{c}{\bf USE (\textuparrow)} 
& \multicolumn{1}{c}{\bf PPL (\textdownarrow)} 
& \multicolumn{1}{c|}{\bf $\Delta$ I (\textdownarrow)} 
& \multicolumn{1}{c}{\bf USE (\textuparrow)} 
& \multicolumn{1}{c}{\bf PPL (\textdownarrow)}
& \multicolumn{1}{c|}{\bf $\Delta$ I (\textdownarrow)}  
& \multicolumn{1}{c}{\bf USE (\textuparrow)} 
& \multicolumn{1}{c}{\bf PPL (\textdownarrow)}
& \multicolumn{1}{c}{\bf $\Delta$ I (\textdownarrow)}  \\

\midrule
            
PWWS
& 0.81 & 812.53 & 2.16
& 0.90 & 1038.79 & 10.24 
& 0.89 & 127.56 & 13.05 \\ 

BERT-ATTACK
& 0.72 & 1257.72 & 0.76
& 0.72 & 889.67 & 0.53
& 0.80 & 2425.45 & 1.00 \\ 

CT-GAT
& 0.65 & 73.62 & 2.84
& 0.81 & 26.66 & -0.08
& 0.88 & 18.90 & 0.05 \\

Q-faker (ours)
& 0.58 & 142.89 & 0.61
& 0.72 & 86.89 & -6.73
& 0.80 & 31.09 & -9.3 \\

\bottomrule
\end{tabular}

\caption{Comparison of generated adversarial examples by attack methods on additional dataset. 
}
\label{appendix_table:result_quality}
\end{table*} 

\subsection{ChatGPT Prompt Evaluation}
We utilized GPT-rank for the evaluation prompt. 
To ensure a fair comparison, we select 166 cases that were successfully attacked by all three methods, BERT-Attack and Q-faker. We use \texttt{GPT-4o-mini} API for evaluation.
Our method outperforms the BERT-Attack in all metrics as shown in Table \ref{appendix_table:chatgpt_eval}

\begin{table}[t]
\small
    \centering
    \begin{minipage}{\linewidth}
        \caption*{Dataset: EDENCE}
        \begin{tabular}{p{7.3cm}}
            \toprule
            \textbf{Original Sentence}  \\
            both frevert and whalley were part of enrons office of the chairman.\\
            \midrule
            \midrule
            \textbf{BERT-Attack} \\
            both fr\tr{ucg} and wha\tr{l}ey were \tr{joint} of enrons office of the chairman.\\
            \midrule
            \textbf{CT-GAT} \\
            \tr{B}oth frev\tr{3}rt \tr{ad} wh\tr{o}alley \tr{we re} \tr{pa rt} of enrons of\tr{if}ce of \tr{th e} chairman.\\
            \midrule
            \textbf{Q-faker (our)} \\
            both frevert and whalley were part of the \tr{same enrons group.}\\
            \bottomrule
        \end{tabular}
    \end{minipage}%
    \caption{
    Case example of an adversarial example generated by attack methods. 
    } 
    \label{appendix_tab:case1}
\end{table}

\subsection{Time-Complexity}
\label{subsection:runtime}
Black-box attacks obtain output information (logit scores or predicted labels) from the target model and use this information to iteratively select substitute words and optimize the order of substitution positions. The computational complexity of these processes is usually worse than $O(n)$, where $n$ is the length of the input sentence.
The majority of black-box attack method has a complexity $O(n^2)$ for finding substitute words and $O(n)$ for optimizing the order, resulting in an overall complexity is $O(n^2 + n)$.
Our proposed method generates half of the sentence without considering which words to substitute or their order, resulting in a complexity $O(1/2 * n)$. Even the complexity can be simplified to $O(1)$ from the perspective of the target model, as our method leverages the gradient information of the surrogate model and never uses the target model's output information. Thus, our method is significantly more effective, with much lower computational complexity than the black-box attacks. 

\subsection{Qualitative Example}
Table \ref{appendix_tab:case1} shows adversarial examples generated by attack methods. 
These examples are the original sentence and crafted adversarial examples in the EDENCE dataset. The results show that the baselines often cause unnatural fluency and grammatical errors in the original sentences. BERT-Attack change important words as name (frevert $\xrightarrow{}$ frucg), this is crucial problem to preserve meaning of sentence. In the case of CT-GAT, the generated text becomes difficult for humans to read. This makes it easy to detect manipulation.
In contrast, our method preserves the meaning of the sentences while maintaining high fluency and grammatical correctness.

\begin{table}[t]
\centering
\small
\begin{tabular}{l|ccc}
\toprule
{\bf Method} & Precision (\textdownarrow) & Recall (\textdownarrow) & F1-score (\textdownarrow)   \\
\midrule
TextFlooler & 58.5 & 57.0 & 55.0  \\
BERT-ATT. & 82.4 & 82.2 & 82.1 \\
CT-GAT & 57.8 & 56.3 & 54.2 \\
Q-faker (our) & \textbf{55.8} & \textbf{54.5} & \textbf{51.8} \\
\bottomrule
\end{tabular}
\caption{Performance with adversarial example detectors: 
A lower score indicates that the detector has been successfully bypassed. 
}
\label{table:defense}
\end{table}

\subsection{Adversarial Defense}
\label{appendix_subsection:defense}
Existing defenses against adversarial attacks encompass various methodologies. 
Broadly, these can be categorized into adversarial training through additional train datasets \citep{dong2021should, zhao2023generative}, and adversarial detection, which aims to detect whether inputs are adversarial examples \citep{mosca-etal-2022-suspicious, zheng-etal-2023-detecting}. 
In real-world scenarios, adversarial detection is more practical.
We conduct experiments using adversarial detector proposed by~\citep{mosca-etal-2022-suspicious}.
The results show strong performance of our method as shown in Table \ref{table:defense}.

\begin{table*}[t]
\centering
\small
\begin{tabular}{c|cc|cc|cc|cc}
\toprule

\multirow{2}{*}{\diagbox{Method}{LLMs}}
& \multicolumn{2}{c|}{LLaMa-7B} 
& \multicolumn{2}{c|}{Mistral-7B} 
& \multicolumn{2}{c|}{DeepSeek-7B} 
& \multicolumn{2}{c}{Gemma2-9B} \\
\cline{2-9}

& \multicolumn{1}{c}{Assassin}
& \multicolumn{1}{c|}{ HSOL} 
& \multicolumn{1}{c}{ Assassin} 
& \multicolumn{1}{c|}{ HSOL} 
& \multicolumn{1}{c}{ Assassin} 
& \multicolumn{1}{c|}{ HSOL} 
& \multicolumn{1}{c}{ Assassin} 
& \multicolumn{1}{c}{ HSOL}  \\

\midrule
CT-GAT & 1.0 & 8.5 & 4.9 & 7.6 & 3.2 & 7.1 & 0.3 & 3.2 \\
Q-faker (ours) & \textbf{2.9} & \textbf{42.5} & \textbf{13.0} & \textbf{43.9} & \textbf{8.0} & \textbf{8.8} & \textbf{11.3} & \textbf{42.8}  \\

\midrule
Clean Acc. & 98.4\% & 97.6\% & 97.9\% & 96.3\% & 63.2\% & 92.2\% & 98.7\% & 96.3\%  \\
\bottomrule
\end{tabular}
\caption{ASR of our method and CT-GAT on various LLMs in zero-shot inference.
}
\label{table:llm_attack}
\end{table*}

\subsection{Feasibility of Attacks on LLMs}
\label{appendix_subsection:feasibility}
Our study focuses on classification language models, rather than generative large language models (LLMs). 
Classification models can be more efficiently utilized in real-world applications, such as automated systems for detecting spam or toxic content. 
Notably, small language models demonstrate classification performance comparable to LLMs while significantly reducing training costs, inference time, and latency.
To validate this, we conducted zero-shot inference on LLMs using 1,000 examples from the Assassin dataset. The fine-tuned BERT model (used as the target model in this study) achieved an accuracy of 98.4\%, whereas Mistral-7B and LLaMA-7B obtained 95.6\% and 98.1\%, respectively. The performance of additional target models is provided in Table \ref{table:target_models}.

\paragraph{LLM Attacks.}
Attacks on LLMs belong to a distinct area of research. Unlike classification language models such as encoder-based BERT, which are primarily used for tasks like text classification, LLMs (e.g., LLaMA-7B) are designed for natural language generation, including conversational responses and question answering. Therefore, adversarial attacks targeting generative models—such as jailbreaking attacks—fall under a separate research domain. A more detailed discussion of these attacks is provided in the Related Work section.

\paragraph{Zero-Shot Inference on Adversarial Examples.}
To assess the feasibility of adversarial attacks on LLMs, we performed zero-shot inference on LLMs using 1,000 adversarial examples generated by CT-GAT and Q-Faker from the Assassin dataset. Table \ref{table:llm_attack} presents ASR results on LLaMA-7B and Mistral-7B. Our experimental findings demonstrate that our method outperforms the baseline and achieves high ASR on LLMs, indicating the feasibility of adversarial attacks on these models.
Our approach leverages pre-trained generative models, which may share knowledge with LLMs, thereby enhancing the attack's effectiveness. 

 
\onecolumn
\section{ChatGPT prompting template - GPT-Rank}
\label{sec:appendix_c}
\begin{table}[h]
    \centering
    \small
        \begin{tabularx}{\linewidth}{l X}
            \toprule
            & \textbf{Instruction}  \\
            & Please read the original text and the two adversarial texts (Candidate-A and Candidate-B), then evaluate and rank texts generated by two different methods. \\
            & \\
            & \textbf{Original Text} \\
            & \{\textit{orig\_text}\} \\
            & \\
            & \textbf{Adversarial Texts} \\
            & \textbf{Candidate-A} : \{\textit{generated\_text1}\} \\
            & \textbf{Candidate-B} : \{\textit{generated\_text2}\} \\
            & \\
            & \textbf{Questions} \\
            Template & Given the instruction and input above, please compare the two candidates based on the \{\textit{metric}\}. \\
            & "\{\textit{metric}\}" \{\textit{metric\_desc}\} \\
            & \\
            & You only have 2 choices to output: \\
            & If you think A is better, please output: 1. Candidate-A is better \\
            & If you think B is better, please output: 2. Candidate-B is better \\
            & \\
            & Do not output anything else except the 2 choices above. \\
            & \\
            & \textbf{Output your choice below Comparison Option (1 or 2)} \\
            & 1. Candidate-A is better \\
            & 2. Candidate-B is better \\
            \midrule
            Variables & \{\textit{orig\_text}\} is original sentence. \\
            &  \\
            & \{\textit{generated\_text}\} is adversarial generated examples by attack methods. \\
            & \\
            &\{\textit{metric}\} is metric to evaluate the quality of generated text. we use three metrics as follows: \\ 
            & \texttt{Naturalness}, \texttt{Meaning Preservation}, \texttt{Grammatical Correctness}.  \\
            & \\
            &  \{\textit{metric\_desc}\} is description of the metric. The description is paired with the following three metrics: \\
            & \texttt{Naturalness} : “evaluates how natural, fluent, and human-like the adversarial example sounds.” \\ & \texttt{Meaning Preservation} : “evaluates whether the original text and the adversarial text have similar meanings.” \\
            & \texttt{Grammatical Correctness} : “checks if the adversarial example is grammatical correct.”\\
            & \\
            \bottomrule
        \end{tabularx}

\caption{GPT-Rank template-based Prompt for evaluation. We utilized GPT-rank for the evaluation prompt. To prevent positional bias, the baseline and our proposed method were randomly assigned to the \{\textit{generated\_text1}\} and \{\textit{generated\_text2}\} positions. 
Additionally, we use "candidate-A" instead of the method's name to avoid naming bias.
To ensure a fair comparison, we select 166 cases that were successfully attacked by all three methods, BERT-Attack, CT-GAT, and Q-faker. We use \texttt{GPT-4o-mini} API for evaluation.}
\end{table}



\end{document}